\documentclass[aps,preprint]{revtex4}
\usepackage{graphicx}
\usepackage[scanall]{psfrag}
\usepackage{amssymb}

\begin{document}

%		DEFINITIONS FOR TEX
%
%%%%%%%%%%%%%%%%%%%%%%%%%%%%%%%%%%%%%%%%%%%%%%%%%%%%%%%%%%%%%%%
%
%
%\def\e{\'e}
%\def\ee{\`e}
%%%%%%%%%%%%%%%%%%%DEFINITIONS%%%%%%%%%%%%%%%%%%%%%%%%%%%%%%%%%
%
\def\oti{{\otimes}}
\def\lb{ \left[ }
\def\rb{ \right]  }
\def\tilde{\widetilde}
\def\bar{\overline}
\def\hat{\widehat}
\def\*{\star}
\def\[{\left[}
\def\]{\right]}
\def\({\left(}		\def\BL{\Bigr(}
\def\){\right)}		\def\BR{\Bigr)}
	\def\BBL{\lb}
	\def\BBR{\rb}
%
%%%%%%%%%%%%%%%%%%%%%%%%%%%%%%%%%%%%%%%%%%%%%%%%%%%%%%%%%%%%%%%
%
\def\zb{{\bar{z} }}
\def\zbar{{\bar{z} }}
\def\frac#1#2{{#1 \over #2}}
\def\inv#1{{1 \over #1}}
\def\half{{1 \over 2}}
\def\d{\partial}
\def\der#1{{\partial \over \partial #1}}
\def\dd#1#2{{\partial #1 \over \partial #2}}
\def\vev#1{\langle #1 \rangle}
\def\ket#1{ | #1 \rangle}
\def\rvac{\hbox{$\vert 0\rangle$}}
\def\lvac{\hbox{$\langle 0 \vert $}}
\def\2pi{\hbox{$2\pi i$}}
\def\e#1{{\rm e}^{^{\textstyle #1}}}
\def\grad#1{\,\nabla\!_{{#1}}\,}
\def\dsl{\raise.15ex\hbox{/}\kern-.57em\partial}
\def\Dsl{\,\raise.15ex\hbox{/}\mkern-.13.5mu D}
%
%%%%%%%%%%%%%%%%%%%%GREEK LETTERS%%%%%%%%%%%%%%%%%%%%%%%%%%%%%%
%
%\def\th{\theta}		\def\Th{\Theta}
\def\ga{\gamma}		\def\Ga{\Gamma}
\def\be{\beta}
\def\al{\alpha}
\def\ep{\epsilon}
\def\vep{\varepsilon}
\def\la{\lambda}	\def\La{\Lambda}
\def\de{\delta}		\def\De{\Delta}
\def\om{\omega}		\def\Om{\Omega}
\def\sig{\sigma}	\def\Sig{\Sigma}
\def\vphi{\varphi}
%
%%%%%%%%%%%%%%%%%%%CALIGRAPHIC LETTERS%%%%%%%%%%%%%%%%%%%%%%%%%
%
\def\CA{{\cal A}}	\def\CB{{\cal B}}	\def\CC{{\cal C}}
\def\CD{{\cal D}}	\def\CE{{\cal E}}	\def\CF{{\cal F}}
\def\CG{{\cal G}}	\def\CH{{\cal H}}	\def\CI{{\cal J}}
\def\CJ{{\cal J}}	\def\CK{{\cal K}}	\def\CL{{\cal L}}
\def\CM{{\cal M}}	\def\CN{{\cal N}}	\def\CO{{\cal O}}
\def\CP{{\cal P}}	\def\CQ{{\cal Q}}	\def\CR{{\cal R}}
\def\CS{{\cal S}}	\def\CT{{\cal T}}	\def\CU{{\cal U}}
\def\CV{{\cal V}}	\def\CW{{\cal W}}	\def\CX{{\cal X}}
\def\CY{{\cal Y}}	\def\CZ{{\cal Z}}

\def\rvac{\hbox{$\vert 0\rangle$}}
\def\lvac{\hbox{$\langle 0 \vert $}}
\def\comm#1#2{ \BBL\ #1\ ,\ #2 \BBR }
\def\2pi{\hbox{$2\pi i$}}
\def\e#1{{\rm e}^{^{\textstyle #1}}}
\def\grad#1{\,\nabla\!_{{#1}}\,}
\def\dsl{\raise.15ex\hbox{/}\kern-.57em\partial}
\def\Dsl{\,\raise.15ex\hbox{/}\mkern-.13.5mu D}
%
%%%%%%%%%%%%%%%%%%%%GREEK LETTERS%%%%%%%%%%%%%%%%%%%%%%%%%%%%%%
%
%%%%%%%%%%%%%%% MATH CHARACTERS %%%%%%%%%%%%%%%%%%%%%%%%%%%%
%
\font\numbers=cmss12
%\font\numbers=cmu10 scaled\magstep1
\font\upright=cmu10 scaled\magstep1
\def\stroke{\vrule height8pt width0.4pt depth-0.1pt}
\def\topfleck{\vrule height8pt width0.5pt depth-5.9pt}
\def\botfleck{\vrule height2pt width0.5pt depth0.1pt}
\def\Zmath{\vcenter{\hbox{\numbers\rlap{\rlap{Z}\kern
0.8pt\topfleck}\kern 2.2pt
                   \rlap Z\kern 6pt\botfleck\kern 1pt}}}
\def\Qmath{\vcenter{\hbox{\upright\rlap{\rlap{Q}\kern
                   3.8pt\stroke}\phantom{Q}}}}
\def\Nmath{\vcenter{\hbox{\upright\rlap{I}\kern 1.7pt N}}}
\def\Cmath{\vcenter{\hbox{\upright\rlap{\rlap{C}\kern
                   3.8pt\stroke}\phantom{C}}}}
\def\Rmath{\vcenter{\hbox{\upright\rlap{I}\kern 1.7pt R}}}
\def\Z{\ifmmode\Zmath\else$\Zmath$\fi}
\def\Q{\ifmmode\Qmath\else$\Qmath$\fi}
\def\N{\ifmmode\Nmath\else$\Nmath$\fi}
\def\C{\ifmmode\Cmath\else$\Cmath$\fi}
\def\R{\ifmmode\Rmath\else$\Rmath$\fi}
%%%%%%%%%%%%%%%%%%%%%%%%%%%%%%%%%%%%%%%%%%%%%%%%%%%%%%%%%%%%%%%%%
 %%%%%%%%%%%%%%%%%% END OF DEFINITIONS %%%%%%%%%%%%%%%%%%%%%%
 %%%%%%%%%%%%%%%%%%%%%%%%%%%%%%%%%%%%%%%%%%%%%%%%%

\def\barray{\begin{eqnarray}}
\def\earray{\end{eqnarray}}
\def\beq{\begin{equation}}
\def\eeq{\end{equation}}

\def\no{\noindent}

\title{Interacting Bose and Fermi gases  in low dimensions 
and the
Riemann hypothesis}
\author{Andr\'e  LeClair}
\affiliation{Newman Laboratory, Cornell University, Ithaca, NY} 
\date{November 2006}

\bigskip\bigskip\bigskip\bigskip

\begin{abstract}

We apply the S-matrix based finite temperature formalism 
 to non-relativistic Bose and Fermi gases in $1+1$
and $2+1$ dimensions.   For the $2+1$ dimensional case
in the constant scattering length approximation, 
the free energy is given in terms of Roger's dilogarithm
in a way analagous to the thermodynamic Bethe ansatz for
the relativistic $1+1$ dimensional case. 
The $1d$ fermionic case with a quasi-periodic 2-body
potential could provide  a physical framework for understanding
the Riemann hypothesis.

\end{abstract}

%\pacs{{\bf ??????}}
%\pacs{{\bf CHECK} 11.10.Hi, 11.55.Ds, 75.10.Jm}

\maketitle

\def\om#1{\omega_{#1}}
\def\Tr{\rm Tr} 
\def\free{\CF} 
\def\xvec{{\bf x}}
\def\kvec{{\bf k}}
\def\kvecp{{\bf k'}}
\def\omk{\om{\kvec}} 
\def\dk{\frac{d^d\kvec}{(2\pi)^d}}
\def\dkind#1{\frac{d^d #1 }{(2\pi)^d}}
\def\dkom{\frac{d^d\kvec}{(2\pi)^d 2 \omk}}
\def\2pid{(2\pi)^d}
\def\fill{{f}}
\def\dkline{\underline{\underline{d\kvec}}}
\def\dklinep{\underline{\underline{d\kvec'}}}
\def\dklineind#1{\underline{\underline{d#1}}}
\def\ket#1{|#1 \rangle}
\def\bra#1{\langle #1 |}
\def\ebH{e^{-\beta H}}
\def\vol{V}
\def\vhat{\hat{v}} 
\def\bfN#1{{\bf{#1}}}
\def\ketbf#1{|{\bf #1}\rangle}
\def\brabf#1{\langle {\bf #1} | }
\def\u{u}
\def\dLR{{{\stackrel{\leftrightarrow}{\d} }}}
\def\Re{{ \it Re}}
\def\Im{{\it Im}}
\def\deltak{\delta^{(\kvec)}}
\def\deltaE{\delta^{(E)}}
\def\n{n}
\def\Fhat{\digamma}
\def\sto{;}
\def\one{{\bf 1}}
\def\Fhatone{\Fhat_1}
\def\CFPI{\CF_{\rm 2PI}}
\def\kernel{{\bf K}}
\def\U{U}
\def\K{\kernel}
\def\dtwo{{\textstyle \frac{d}{2} }}
\def\dplustwo{{ \textstyle \frac{d+2}{2} }}
\def\dplusone{{ \textstyle \frac{d+1}{2}}}
\def\dminustwo{{ \textstyle \frac{d-2}{2} }}
\def\dminusone{{ \textstyle \frac{d-1}{2}}}
\def\teehat{\hat{T}}
\def\xvec{{\bf x}}
\def\kvec{{ \bf k}}
\def\psidag{\psi^\dagger}
\def\zmu{z_\mu}
\def\zdelta{z_\delta}
\def\Li{{\rm Li}}

%DEFINITIONS

\section{Introduction}

Quantum gases at finite temperature and density 
have very wide applications in nearly all areas
of physics, ranging from black body radiation, 
Fermi liquids, Bose-Einstein condensation, 
and equations of state in cosmology.   
If one knows the zero temperature dynamics, i.e. 
the complete spectrum of the hamiltonian, then
the quantum statistical mechanics just requires
an additional statistical summation $Z = Tr (e^{-\beta H})$,
so that in principle zero temperature dynamics and 
quantum statistical sums are decoupled.   
This decoupling is also in principle clear from 
an intuitive picture of a gas as a finite density of
particles that are subject to scattering.   In practice,
the complete non-perturbative spectrum of $H$ is unknown
so one must resort to perturbative methods, such
as the Matsubara approach,  that 
typically entangle the zero temperature dynamics from
the quantum statistical mechanics.    

In \cite{AndreSthermo} an alternative approach to finite
temperature quantum field theory was developed that 
achieves this decoupling of zero temperature dynamics
and quantum statistical summation.  The dynamical variables
are the occupation numbers, or filling fractions, of the gas,
and the zero temperature data of the underlying theory are 
just the S-matrix scattering amplitudes.
Our  construction 
was  modeled after Yang and Yang's thermodynamic Bethe
ansatz (TBA)\cite{YangYang}.    The TBA is very specific to
integrable theories in one spacial dimension.  Nevertheless, our derivation
\cite{AndreSthermo} is carried out in any spacial dimension $d$
and doesn't assume integrability;  therefore it is in general
an approximate method,  though it can be systematically improved.  
An important ingredient of our construction is the work
of Dashen, Ma, and Bernstein\cite{Ma} which explains how 
to formulate quantum statistical mechanics in terms of the S-matrix.  
Ideas of Lee and Yang were also instrumental\cite{LeeYang},
even though the latter work is not based on the S-matrix.

Our previous work\cite{AndreSthermo} was mainly devoted to developing
the formalism in a general, model independent way for both
non-relativitistic and relativistic theories.   Though
the derivation was somewhat involved,  the final result is 
straightforward to implement and is summarized by the two formulas
(\ref{2bod.5}, \ref{2bod.6}).    In this paper
we apply the method to non-relativistic Bose and Fermi gases 
with special attention paid to the low dimensional cases of $d=1,2$. 
The main approximation we make is to consider only the contributions
to the free energy that come from 2-particle to 2-particle scattering.
On physical grounds these are expected to be the most important if
the gas is not too dense.     Even in this approximation the 
problem is non-trivial since one needs to resum all the 2-body
interactions self-consistently.  This is accomplished by
an integral equation for the filling fractions that is
analagous to the TBA equations.

In the next section we first review the main results in \cite{AndreSthermo}.  
In section III we describe the kind of second-quantized hamiltonians
that are the subject of this paper.   In section IV we consider  
bosons with a $\delta$ function two-body potential.
In section V we
turn to fermionic gases in the same approximation
and derive formulas that can be used
to study the effect of interactions on the Fermi energy.
Since the lowest
order contribution corresponds to a constant scattering length approximation,
the results we obtain in sections IV and V are
already known, and are included here mainly for illustrative purposes.    

In section IV we study the $2d$ case which has some remarkable
features.    The integral equation which determines the filling fraction
becomes algebraic.   The formulas for the free energy are
essentially identical to those for the conformally invariant
limit of {\it relativistic} theories in {\it one lower} dimension 
$d=1$\cite{ZamoTBA} and are given in terms of Roger's dilogarithm.  
In making this analogy we define a ``central charge'' $c$ as the
coefficient in the free energy, which in the relativistic case in
$1d$ is the Virasoro central charge.    As in the $1d$ relativistic
case,   for certain special values of the coupling,  $c$ can
be a rational number, and some examples based on the golden mean
are presented.   We also extend the formalism to many species
of mixed bosonic and fermionic statistics.

As will be clear in the sequel,  
the polylogarithmic and Riemann $\zeta$ functions play a central
role in this work.    This provided us with a new opportunity
to understand Riemann's hypothesis\cite{Edwards}.  
   Riemann's $\zeta$ function 
for $\Re (\nu ) > 1$ is defined as 
\beq
\label{Riem.1}
\zeta (\nu) = \sum_{n=1}^\infty \inv{n^\nu} 
\eeq
It has a simple  pole at $\nu = 1$;  as we will describe 
this pole is a manifestation of the impossibility 
of Bose-Einstein condensation in  $2d$.   The function can be analytically
continued to the whole complex $\nu$ plane.   It has trivial
zeros at $\nu = -2n$ where $n$ is a non-zero positive integer. 
The function has a duality relation that relates $\zeta (\nu)$
to $\zeta (1-\nu)$ (eq. (\ref{ap.8}) below) so that its non-trivial zeros are 
in the critical strip $0< \Re (\nu) < 1$ and symmetric around
$\Re (\nu ) = 1/2$.   
The Riemann hypothesis (RH)  is the statement that the only
non-trivial zeros of $\zeta (\nu) $ are at $\Re (\nu) = 1/2$. 
This hypothesis is very important in number theory since,
as shown by Riemann,  the distribution of prime numbers
is intimately related to the location of the Riemann zeros.  
It has already been proven by Hadamard and de la Vall\'ee-Poussin
that there are no zeros with $\Re (\nu) =1$, which is equivalent
to proving the Prime Number Theorem\cite{Edwards}.

There have been a number of approaches to understanding
the Riemann hypothesis based on physics\footnote{For a comprehensive
list see M. Watkins at
http://secamlocal.ex.ac.uk/~mwatkins/zeta/physics.htm.}.  Here we mention 
a few of them mainly to contrast with the approach presented here. 
Some approaches are inspired by the Polya-Hilbert conjecture
which supposes that the imaginary part $\alpha$ of the zeros
$\nu = 1/2 + i \alpha$ corresponds to eigenenergies of
some unknown quantum mechanical hamiltonian.    
The well-studied statistical properties
of the zeros, in particular the Montgomery-Odlyzko law
and its relation to the gaussian unitary ensemble  of random 
matrices\cite{Montgomery,Odlyzko}, led   
Berry to propose  a quantum chaos interpretation 
of the oscillatory part of Riemann's counting formula 
for the zeros\cite{Berry}.  This is  still within the context of 
quantum mechanics at zero temperature.
Connes has proposed the complementary picture
that the
zeros correspond to absorption lines\cite{Connes}.    
Sierra has recently proposed a consistent quantization of
the Berry-Keating hamiltonian based on the Russian doll model
of superconductivity, which possesses a cyclic renormalization
group flow\cite{Sierra}.

The fundamental duality relation relating $\zeta(\nu)$ to
$\zeta (1-\nu)$ can be understood as a consequence of
a special modular transformation for the quantum statistical
mechanics of free, relativistic, massless particles in 
$\nu = d+1$ spacetime dimensions.   This is explained in
the appendix,  since it is tangential to the main development of
this paper.       
In section VII   Riemann's
$\zeta$ function on the critical strip
is  related to the quantum statistical mechanics
of {\it non-relativistic}, 
interacting {\it fermionic} gases in $1d$ with a quasi-periodic
2-body potential which depends on $\nu$. This is thus
a many-body problem at finite temperature.    It is perhaps not
completely unanticipated that the RH could have a resolution in
the present context, since one of the very first places in
physics where $\zeta$ appeared was in Planck's work on black body 
radiation,  which  is widely acknowledged as the birth of quantum 
mechanics\cite{Planck}. 
The black body theory is 
the bosonic quantum statistical mechanics of a $3d$ gas of free relativistic
photons\footnote{It is not clear
from Planck's paper whether he was aware he
was dealing with Riemann's $\zeta$ function.   
He simply writes $1 + \inv{2^4} + \inv{3^4} + \inv{4^4} + \cdots  =
1.0823$}.  
In this case what appears is $\zeta (4) = \pi^4 /90$.  
As we describe here,  in order to get into the critical strip
one needs $d=1$ non-relativistic fermions interacting with a quasi-periodic 
potential.    The fermionic nature renders the required integrals
convergent and is also needed for a well-defined hamiltonian.   
This quasi-periodicity implies our approach is closest to
Sierra's\cite{Sierra}, 
since there also a periodicity was
important, but in the renormalization group at zero temperature. 
Our approach is thus essentially different just because
the physical context is different, however  
our interest in the RH actually stemmed from the work\cite{RD} 
on cyclic renormalization group flows in $1d$ relativistic
systems and the observation that the finite temperature 
behavior was in part characterized by $\zeta (1 - i \alpha)$,
where $2\pi/\alpha$ is the period of the renormalization group flow\cite{RD2}.
We now understand that one needs to consider non-relativistic
models to get $\Re (\nu) < 1$ in a natural way\footnote{There 
are also some convergence problems with the usual TBA, which 
are avoided in the present work.}.

The understanding of the RH that emerges from the present context 
can be summarized as follows.   When $\Re (\nu) > 1/2$, 
there exists a two-body potential in position space that leads 
to a well-defined quasi-perodic kernel $\K$  in momentum space 
that is essentially a combination of Fourier transforms of the
potential.    The condition $\Re (\nu) > 1/2$ comes about
naturally as the condition for the convergence of
the appropriate Fourier transform. 
The quantum statistical mechanics based on this
kernel gives corrections to the pressure of the gas that are 
determined by a transcendental equation involving the
polylogarithm $\Li_\nu$.   
To obtain this, one  must work with the momentum-dependent kernel,
i.e. the constant scattering length approximation made in
earlier sections vanishes.  
If $\zeta (\nu ) = 0$ then
there would exist solutions with vanishing corrections to the 
pressure.    Therefore the RH would follow from the simple physical
property that non-zero interactions necessarily modify the pressure. 
We comment on the meaning of the actual zeros in section VII.

\section{Free energy as a dynamical functional of filling fraction}

\subsection{Generalities} 

The free energy density (per volume) $\CF$ is defined as 
\beq
\label{L1}
\CF = - \inv{\beta V} \log Z , ~~~~~~~ Z = \Tr ~ e^{-\beta (H-\mu N)} 
\eeq
where $\beta = 1/T, \mu $ are  the inverse temperature and chemical
potential, $V$ is the d-dimensional spacial volume,  and $H$ and $N$
are the hamiltonian and particle number operator.  Since 
$\log Z$ is an extensive quantity, i.e. proportional to the volume, 
the pressure $p$ of the gas is minus the free energy density:
\beq
\label{pressure}
p= T \frac{d \log Z}{dV} = - \CF
\eeq
For most of this paper, 
we  assume there is one species of bosonic ($s=1$) or
fermionic ($s=-1$) particle.   Given $\CF (\mu)$, one can compute
the thermally averaged number density $\n$:
\beq
\label{L2}
n = - \frac{\partial \CF}{\partial \mu } \equiv  
\int \dk ~ f(\kvec)
\eeq 
where $\kvec$ is the d-dimensional momentum.
The dimensionless
quantities $f$ are called  the filling fractions or
occupation numbers.      

One can express $\CF$ as a functional of $f$ in a meaningful
way with a Legendre transformation.  Define
\beq
\label{L3}
G \equiv \CF (\mu) + \mu \,  n  
\eeq
Treating $f$ and $\mu$ as independent variables,  then 
using eq. (\ref{L2}) one has that 
$\d_\mu G =0$ which implies it can be expressed only in terms of 
$f$ and satisfies $\delta G/ \delta f = \mu$.   

Inverting the above construction shows that there exists 
a functional $\Fhat (f, \mu)$ 
\beq
\label{L4}
\Fhat (f, \mu) = G(f) - \mu \int \dk ~ f(\kvec) 
\eeq
which satisfies eq. (\ref{L2}) and is a stationary point 
with respect to $f$:
\beq
\label{L5}
\frac{\delta \Fhat}{\delta f} = 0
\eeq
The above stationary condition 
 is to be viewed as determining
$f$ as a function of $\mu$.   The physical free energy is
then $\CF = \Fhat$ evaluated at the solution $f$ to the above equation.
We will refer to eq. (\ref{L5}) as the saddle point equation since
it is suggestive of a saddle point approximation to 
a functional integral:
\beq
\label{saddle}
Z = \int D f  ~ e^{-\beta V \Fhat (f)} \approx  e^{-\beta V \CF} 
\eeq

In a free theory,  the eigenstates of $H$ are multi-particle
Fock space states $|\kvec_1 , \kvec_2 ....\rangle$. 
Let  $\omega_\kvec$ denote the 
one-particle energy as a function of momentum $\kvec$. 
In this paper the theory is assumed to be non-relativistic
with 
\beq
\label{omegadef}
\omega_\kvec = \frac{\kvec^2}{2m} 
\eeq
where $m$ is the mass of the
particle.   
It is well-known that the trace over the multi-particle 
Fock space gives
\beq
\label{L6}
\CF_0 (\mu) = \frac{s}{\beta} \int \dk ~ 
\log \( 1 - s e^{-\beta (\om\kvec - \mu )} \) 
\eeq
From the definition eq. (\ref{L2}) one finds the filling fractions: 
\beq
\label{L7}
f (\kvec) = \inv{ e^{\beta(\om\kvec - \mu)} -s } \equiv f_0 (\kvec) 
\eeq
In order to find the functional $\Fhat (f,\mu)$ one first computes
$G$ from eq. (\ref{L3}) and eliminates $\mu$ to express it in terms 
of $f$ using eq. (\ref{L7}).  One finds
\beq
\label{L8}
\Fhat_0 (f,\mu) = \int \dk \( 
(\om\kvec - \mu ) f - \inv\beta 
\Bigl[ (f+s) \log (1 + sf) - f \log f \Bigr] \) 
\eeq
One can then easily verify that $\delta \Fhat / \delta f = 0$
has the solution  $f=f_0$ and plugging this back into eq. (\ref{L8}) gives
the correct result eq. (\ref{L6})  for $\CF_0$.  
In the sequel, it will be convenient to trade the chemical
potential variable $\mu$ for the variable $f_0$:  
\beq
\label{L10}
\Fhat_0 (f, f_0) = -\inv{\beta} \int \dk 
\(  s \log (1 + sf)  +  f 
\log \( \frac{1+sf}{f} \frac{f_0}{1+s f_0} \)  \)
\eeq

There is another way to view the above construction which 
involves the entropy.  Write eq. (\ref{L8}) as 
\beq
\label{L9}
\Fhat = \CE - \inv\beta \,  \CS
\eeq
where $\CE$ is the first $(\omega -\mu)f$ term in eq. (\ref{L8}), 
which is the energy density,  
and $\CS$ is the remaining term in brackets.   
One can show by a standard counting argument, which  involves 
the statistics of the particles,  that $\CS$ 
represents the entropy density  of a gas of particles.  
(See for instance \cite{Landau}.)

Let us now include interactions by writing 
\beq
\label{var.1}
\Fhat (f,f_0) = \Fhat_0 (f,f_0) + \Fhatone (f) 
\eeq
where $\Fhat_0$ is given in eq. (\ref{L10}) and we define
$\U$ as the ``potential'' which depends on $f$ and 
incorporates interactions:
\beq
\label{var.2}
\Fhatone = -\inv{\beta} 
\int \dkind\kvec ~ \U (f(\kvec) )
\eeq
Given $\Fhat$,  $f$ is determined by the saddle point equation
$\delta \Fhat / \delta f = 0$.  
It is convenient to define a pseudo-energy $\vep$ as the
following parameterization of $f$:
\beq
\label{pseudo.1}
f \equiv \inv{ e^{\beta \vep} -s }
\eeq
Then the saddle point equation and free energy density take the form:
\barray
\label{pseudo.2}
\vep &=& \omega - \mu - \inv{\beta} \frac{\d \U}{\d f} 
\\ 
\label{pseudo.3}
\CF &=& - \inv{\beta} \int  \dkind{\kvec} 
\[ -s \log (1 - s e^{-\beta \vep} ) 
+ (1- f\d_f  ) \U  \]
\earray

\subsection{Two-body approximation}

It was shown in \cite{AndreSthermo} how to express $\U$ in terms
of S-matrix scattering amplitudes.  In general there are terms
involving the forward scattering of $n \to n$ numbers of particles
for $n\geq 2$.     The most important is the two-body contribution.
It takes the form  
\beq
\label{2bod.2}
\U (\kvec ) = \frac\beta {2}  f(\kvec )  ~  (\K * f)(\kvec )  
\eeq
where we have defined the convolution
\beq
\label{2bodstar}
(\K * f )(\kvec) \equiv  \int \dkind{\kvec'} ~ \K (\kvec, \kvec' ) f(\kvec'), 
\eeq
and the kernel $\K$ is given by the 2-particle to 2-particle
forward scattering amplitude:
\beq
\label{2bod.1}
\K (\kvec , \kvec' ) \equiv  
\CM_{\bf 12;12} (\kvec, \kvec' ) 
\eeq
How to compute the kernel from  the hamiltonian is described below.

In terms of the pseudo-energy,  the saddle point equation 
and free energy take the following forms:
\barray
\label{2bod.5}
\vep &=& \omega - \mu -  \K * \( \inv{e^{\beta \vep} -s } \)
\\
\label{2bod.6}
\CF &=&
-\inv{\beta} \int \dkind\kvec 
\[ -s \log (1-s e^{-\beta \vep}) + \frac{\beta f}{2} (\vep - \omega + \mu) \]
\earray

\def\rvec{{\bf x}}
\def\xvec{{ \bf x}}
\def\adag{a^\dagger}

\section{Second quantized hamiltonians}

In this paper we consider second quantized hamiltonians of
the form:
\beq
\label{Ham.1}
H = \int  d^d \rvec 
\({  \inv{2m}} 
 \vec{\nabla} \psi^\dagger \cdot \vec{\nabla} \psi \)  + 
{  \inv{4}} \int d^d  \xvec \int d^d \xvec'  ~ \CV (\xvec - \xvec') 
~ \psidag (\xvec ) \psidag (\xvec' ) 
\psi (\xvec' ) \psi (\xvec ) 
\eeq
where $\CV$ is the 2-body potential.  
The field satisfies the canonical commutation relations
\beq
\label{ham.2}
 \psi (\xvec ) \psidag (\xvec') 
- s\, \psidag (\xvec' ) \psi (\xvec)   = \delta(\xvec - \xvec' ) 
\eeq
where again  $s=\pm 1$ corresponds to bosons/fermions. 
Expanding the field in terms of annihilation operators, 
\beq
\label{nonrel.7}
\psi (\rvec ) = \int \dkind\kvec ~ e^{i\kvec \cdot \rvec} ~ a_\kvec 
\eeq
this leads to the canonical (anti-) commutation relations:
\beq
\label{ham.3}
   a_\kvec   \adag_{\kvec'} - s \, \adag_{\kvec'} a_\kvec  
= (2\pi)^d \delta( \kvec - \kvec' )
\eeq
The free Hilbert space is thus a bosonic or fermionic Fock space 
with  momentum eigenstates  normalized as:
$| \kvec_1 , \kvec_2 , ..., \kvec_n \rangle = \adag_{\kvec_1} 
\cdots \adag_{\kvec_n} | 0 \rangle$.  

In this paper we only consider the lowest order contribution to
the kernel $\K$.   Let $H_1$ denote the interacting part of
$H$ which depends on the 2-body potential $\CV$.  
Then to lowest order,    
\beq
\label{nonrel.6}
\langle \alpha' | H_1 | \alpha \rangle = - (2\pi)^d \delta^{(d)} 
(\kvec_\alpha - \kvec_{\alpha'})
~ \CM_{\alpha' ; \alpha}
\eeq
where $\CM$ is the scattering amplitude for the asymptotic states
$|\alpha \rangle $.    
Therefore  to lowest order the kernel is given by
\beq
\label{nonrel.8}
\K (\kvec , \kvec' ) = - V^{-1} ~ 
\langle \kvec , \kvec' | H_1 | \kvec, \kvec' \rangle 
\eeq

\def\coupling{\gamma}

\section{Hard core boson model revisited}

\def\rvec{{\bf x}}

In this section we 
consider bosonic particles with a delta-function  two-body potential:
\beq
\label{nonrel.4}
\CV (\rvec , \rvec' ) = \coupling \, \delta^{(d)} (\rvec - \rvec' )
\eeq
The model has been solved exactly in $1d$ by Lieb and Liniger\cite{Lieb},
and the thermodynamic Bethe ansatz was first discovered in the 
context of this model\cite{YangYang}.

Using $(2\pi)^d \delta^{(d)} (0) = V$,   one finds
\beq
\label{nonrel.8b}
\kernel  = -\coupling  
\eeq
Positive $\gamma$ corresponds to repulsive interactions. 
Since the kernel is constant in this approximation, it is
equivalent to the constant scattering length approximation
made in the literature; hence we do not obtain any new results
for the hard core Bose gas here.  We include this section
and the next mainly for illustrative purposes and as a warm-up
to the sequel.

The coupling constant $\coupling$ has units of 
${\rm energy} \times {\rm volume}$.  It can be expressed in 
terms of a physical scattering length $a$ as follows. 
To first order in perturbation theory the differential 
cross-section in the center of mass is:
\beq
\label{nonrel.9}
\frac{ d\sigma}{d\Omega} = \frac{m^2 \coupling^2}{4 (2\pi)^{d-1} } ~k^{d-3} 
\eeq
where $k$ is the magnitude of $\kvec$ for one of the incoming particles.  
Since a cross section has dimensions of ${\rm length}^{d-1}$,  we define
$a$ such that the cross section is $a^{d-1}$ when the wavelength of the
particle is $2\pi/a$: 
\beq
\label{nonrel.10}
\frac{ d\sigma}{d\Omega} \Bigr\vert_{k \sim 2\pi /a} \sim a^{d-1} 
\eeq
This leads us to make  the definition:
\beq
\label{nonrel.11}
\frac{\coupling}{(2\pi)^{d/2}} \equiv  \frac{a^{d-2}}{m}
\eeq

\def\hT{h_T}
\def\zdelta{{z_\delta}}
\def\Li{{\rm Li}}
\def\lambdaT{\lambda_T}

We carry out our analysis for arbitrary 
spacial dimension $d>0$. 
Using $\int d^d \kvec = \frac{2 \pi^{d/2}}{  \Gamma (d/2)} \int dk ~ k^{d-1}$ 
where $\Gamma$ is the standard $\Gamma$-function,  
rotationally invariant 
integrals over momenta $\kvec$  can  be traded for integrals over $\omega_\kvec$:
\beq
\label{nonrel.12}
\int  \dkind\kvec  =  \( \frac{m}{2\pi} \)^{d/2} \inv{\Gamma(d/2)} 
 \int_0^\infty d\omega 
~ \omega^{(d-2)/2}
\eeq

For a constant kernel $\kernel = -\coupling$, and $d>0$,  the solution to the integral 
equation eq. (\ref{2bod.5}) takes the simple form: 
\beq
\label{nonrel.13}
\vep (\kvec) = \om\kvec - \mu + T \delta
\eeq
where
$\delta$ is independent of $\kvec$.
To determine $\delta$ one needs the integral: 
\beq
\label{bosonint} 
\int_0^\infty dx  ~ \frac{ z \, x^{\nu-1}}{ e^x  -z } = \Gamma (\nu) \Li_\nu (z) ~~~~~~
  {\rm for} ~ \Re (\nu)  >0
\eeq
The function $\Li_\nu (z)$ is the standard polylogarithm, defined as the appropriate
analytic continuation of
\beq
\label{nonrel.17}
\Li_\nu (z) = \sum_{n=1}^\infty \frac{z^n}{n^\nu}
\eeq
For fixed $\nu$, the function $\Li_\nu (z)$ has a branch point at $z=1$ 
with a cut along $\Re (z) > 1$, 
and the above integral  is not valid at $z=1$.    
We will later need the fact that when $z=1$,  the above integral is only convergent 
when $\Re (\nu)  > 1$ and is given in terms of Riemann's $\zeta$ function:
\beq
\label{zetaintboson}
\int_0^\infty  dx ~ \frac{x^{\nu -1}}{e^x -1 } = \Gamma (\nu) \zeta (\nu), ~~~~~\Re (\nu) > 1 
\eeq 
Equivalently, 
\beq
\label{lizeta}
\Li_{\nu} (1) = \zeta(\nu) , ~~~~~~~\Re (\nu) > 1 
\eeq

Eq. (\ref{2bod.5}) then  leads to the following  equation satisfied by $\delta$: 
\beq
\label{nonrel.14}
\delta =  \hT  ~  \Li_{d/2} ( z_\mu \zdelta )
\eeq
where we have defined  the fugacities 
\beq
\label{nonrel.1}
 z_\mu  \equiv e^{\beta \mu} , ~~~~~ \zdelta = e^{-\delta} 
\eeq
and a renormalized thermal coupling $\hT$ and thermal wavelength $\lambdaT$:
\beq
\label{nonrel.16}
\hT  \equiv \( \frac{\sqrt{2\pi}a}{\lambdaT} \)^{d-2},  
~~~~~ \lambdaT  \equiv \sqrt{ \frac{2\pi}{mT} } 
\eeq

\def\Ttilde{{\tilde{T}}}

To simplify subsequent  expressions, we henceforth set the mass $m=1/2$,
unless otherwise stated.   This leads us to define:
\beq
\label{Ttildeeq}
\tilde{T} \equiv \inv{\lambdaT^2} = 
  \frac{T}{4\pi} , ~~~~~(m=1/2)
\eeq

The equation (\ref{nonrel.14}) is a  transcendental  equation 
that determines $\delta$ as a function of $\mu, T$, and the coupling $\hT$. 
Given the solution $\delta (\mu )$ of this equation, using 
eq. (\ref{2bod.6}) the density  can be expressed as
\beq
\label{nonrel.18}
n(\mu )  =  \Ttilde^{d/2} ~ 
\Li_{d/2} ( z_\mu \zdelta  ) 
\eeq
For $d>0$ one can integrate by parts and obtain the following expressions
for the free energy 
\beq
\label{nonrel.18b}
\CF = - T ~  \Ttilde^{d/2} 
\( \Li_{(d+2)/2 } ( z_\mu \zdelta )  + 
\frac{\delta }{2} \Li_{d/2} ( z_\mu \zdelta  ) \) 
\eeq

\def\nphys{n_{\rm phys.}}
\def\dtwo{{\textstyle \frac{d}{2} }}
\def\dplus{{ \textstyle \frac{d+2}{2} }}

Bose-Einstein condensation can be described rather generally as follows.   
The property that an extensive,  i.e. proportional to the volume, 
 number of particles occupy the ground
state with $|\kvec| =0$ implies that the filling fraction $f$ diverges
at some critical chemical potential $\mu_c$:
\beq
\label{BEC.1}
\lim_{\mu \to \mu_c } ~ f(|\kvec|=0, \mu) = \infty
\eeq
Given $\mu_c$, one can speak of a critical density $n_c$:
\beq
\label{BEC.2}
n_c (T) = \int \dkind\kvec ~ f(\kvec , \mu_c ) 
\eeq
One also has a  critical temperature $T_c$ 
defined as 
\beq
\label{BEC.3}
n_c (T_c) = \nphys
\eeq
where $\nphys$ is the physical density. 
The reason that $T_c$ corresponds to a transition is that because
of the divergence, one must treat the density of particles in the
ground state $n_{gs}$ separately from the above expressions for $n$: 
\beq
\label{BEC.4}
\nphys = n_{gs} + n 
\eeq
where $n_{gs}$ is the density of particles in the ground state and 
$n$ is the integral of $f$.  
When $T>T_c$,  $n_{gs}$ vanishes.  The number of particles in the ground
state doesn't actually vanish, but rather it is no longer proportional
to the volume, so that $n_{gs}$ effectively vanishes.

\def\dplustwo{{\textstyle \frac{d+2}{2} }}

In the pseudo-energy description (\ref{pseudo.1}) of the filling fraction $f$, 
the condition (\ref{BEC.1}) is just 
\beq
\label{BEC.4b}
\vep(|\kvec| =0, \mu = \mu_c ) = 0
\eeq 
Eq. (\ref{2bod.5})  then leads to the equation for $\mu_c$: 
$\mu_c = T \delta (\mu_c)$. 
In terms of the fugacities the critical point is the simple relation
\beq
\label{fugcrit}
z_{\mu_c}   z_{\delta (\mu_c )}  =1
\eeq
At the critical point,  the arguments of the polylogarithms  are $1$,  and
if $d>2$,  then 
$\Li_{d/2} (1) = \zeta (d/2  )$.  (Eq. (\ref{lizeta})).     
This in turn implies that the critical temperature and density are independent of 
the coupling $\hT$:
\barray
\label{critmu}
\mu_c &=&   \hT \, \zeta(d/2) 
\\ 
\label{critden}
n_c &=& \zeta (d/2)   \Ttilde^{d/2} 
\\
\label{crittemp}
T_c &=& 4\pi  \(  \frac{\nphys}{\zeta(d/2)} \)^{2/d}  
\earray 
Though the critical temperature  and density 
do not depend on the coupling $\hT$ to
this order,  the
free energy at the critical point has corrections that do:
\beq
\label{BEC.6}
\CF = -  \zeta(\dplustwo  )\,  T ~   \Ttilde^{d/2}  
\( 1 + \frac{\hT}{2}    \frac{\zeta(\dtwo )^2 }{\zeta(\dplustwo)} \)   
\eeq

\def\2pilambda{2\pi \lambdaT^2}

The $\zeta (\nu) $ function has a simple pole at $\nu = 1$,
and this has a physical significance in the present context.  
In particular, it implies that the critical density $n_c$ is infinite and the
critical temperature $T_c$
is zero  in $2d$ to the order we have calculated.   
This means that in $3d$ Bose particles have a  stronger  tendency
to Bose-Einstein condense in comparison with $2d$.
This simple pole at $\nu =1$ can thus be interpreted as a
manifestation of the Mermin-Wagner theorem, which states that
finite temperature continuous phase transitions are not possible
in $2d$\cite{Mermin}. 
Another way of viewing this is that in $2d$,  bosonic particles behave
more like fermions and don't readily Bose-Einstein condense.    
This fact is ultimately responsible for why one needs to treat
bosonic particles with fermionic exclusion statistics in $1d$
as far as their quantum statistical mechanics is concerned.  
See for instance \cite{Gurarie}.  For relativistic models also,
it appears that only the fermionic thermodynamic Bethe ansatz equations
are consistent\cite{Klassen,Mussardo}.

\section{Fermi gases  and the Fermi energy}

In this section we consider fermionic particles 
with the same constant kernel $\K = - \gamma$ as in the last section.    
To obtain such a kernel from the second quantized hamiltonian
one needs at least two species of fermions since $\psi^2 = 0$, 
however for simplicity we consider only one species.   
 
The integral we  need is now:
\beq
\label{fermint}
\int_0^\infty dx ~ \frac{z\, x^{\nu -1}}{e^x  + z} = - 
\Gamma (\nu) \Li_{\nu} (-z) 
~~~~~~\Re (\nu ) > 0
\eeq
It is important to realize that contrary to the bosonic case,
eq. (\ref{zetaintboson}),  the above integral is valid when 
$z=1$ throughout the critical strip $0< \Re (\nu) < 1$ and
is given by the above formula where:
\beq
\label{fermzeta}
- \Li_\nu (-1) = (1-2^{1-\nu}) \zeta (\nu) 
~~~~~~ \forall ~ \nu 
\eeq
In physical terms,  fermions are much more stable than bosons  in $d<2$ 
because of  the Pauli exclusion principle.

The pseudo-energy still takes the form (\ref{nonrel.13}), and 
the saddle point equation (\ref{2bod.5}) now leads to the 
equation:
\beq
\label{ferm.1}
\delta = - \hT  \, \Li_{d/2} (-\zmu \zdelta) 
\eeq
The density and free energy can be expressed as: 
\barray
\label{ferm.2}
n &=& - \Ttilde^{d/2} ~   ~ \Li_{d/2} ( - \zmu \zdelta ) 
\\ \nonumber
\CF &=& T ~   \Ttilde^{d/2}   \Bigl( \Li_{(d+2)/2} 
( -\zmu \zdelta ) 
+ \frac{\delta}{2} \Li_{d/2} ( -\zmu \zdelta ) \Bigr) 
\earray

\def\omegaF{{\omega_F}}
\def\kF{{ \kvec_F }}

In the fermionic case, the considerations of the last section
are replaced by the concept of a Fermi surface.  As we'll see,
mathematically there are analogies between the existence of the
Fermi surface and Bose-Einstein condensation.       
The Fermi energy $\omegaF$ is defined as the uppermost
energy that is occupied, i.e. has $f \neq 0$.   In the limit
of zero temperature and no interactions, $f$ is a step 
function:  $f(\kvec) = 1$ for $|\kvec| < |\kF|$ and 
zero otherwise, where $\omegaF = \kF^2 / 2m$.    At finite
temperature the sharp step is broadened.      

The formulas eqs. (\ref{ferm.1},\ref{ferm.2}) can be used
to study the effect of interactions and finite temperature
on $\omegaF$.    The Fermi energy naturally can be defined
as the point where $f = 1/2$, i.e. where $\vep =0$:
\beq
\label{fermienergy.1}
\vep( \kF ) = 0  = \omega_F - \mu + T\delta
\eeq
The latter implies: 
\beq
\label{fermienergy.2}
\zmu \zdelta = e^{\beta \omega_F}
\eeq
The Fermi energy may now be expressed in terms of the density $n$:
\beq
\label{fermienergy.3}
n = - \Ttilde^{d/2}  ~ \Li_{d/2} ( - e^{\beta \omega_F})
\eeq

The above formula determines $\omegaF$ as a function of
$T$ and $n$.   As the temperature goes to zero, one can use
\beq
\label{fermienergy.4}
\lim_{T\to 0} \Li_{\nu} ( - e^{\omegaF /T} ) \approx  
-\frac{ (\omegaF /T)^\nu }{\Gamma (\nu +1 )} 
\eeq
to obtain 
\beq
\label{fermienergy.5}
\omegaF = 4\pi  
\( \Gamma ( \dplustwo ) n \)^{2/d}
\eeq
In $3d$ the above formula is equivalent to well-known results\cite{Landau}. 

Note that, to the order we have done the computation,  the
Fermi energy  does not depend on the interactions since
eq. (\ref{fermienergy.3}) no longer depends on the interaction.   
However in the presence of interactions $\delta \neq 0$,
the free energy is modified from its non-interacting value, 
as eq. (\ref{ferm.2}) shows.

\section{$2d$ Bose and Fermi gases}

In this section we consider the $2d$ case which is rather special. 
Refering to eq. (\ref{nonrel.16}) the coupling $\hT$ is temperature
independent in $2d$, and is in fact dimensionless.  This implies
that in  $2d$ the
scattering length $a^{d-2}$ should simply be replaced by a 
dimensionless coupling we will denote as $h$.    
Many of the
previous formulas are especially simple due to the identity
$ \Li_1 (z) = - \log (1-z) $.    We also describe a 
strong similarity to the conformally invariant limit
of relativistic systems in one lower  dimension.

\subsection{Bosonic case}

The equation (\ref{nonrel.14}) for $\delta$ becomes algebraic: 
\beq
\label{2dboson.1}
\zdelta = (1-\zmu \zdelta)^h
\eeq
The density is simply a logarithm 
\beq
\label{2dboson.2}
n = - \frac{T}{4\pi}   \log (1 - \zmu \zdelta ) 
\eeq
and the free energy is expressed in terms of the dilogarithm:
\beq
\label{2dboson.3}
\CF = - \frac{T^2}{4\pi}   \( \Li_2  (\zmu \zdelta ) - \frac{\delta}{2} 
\log( 1-\zmu \zdelta ) \)
\eeq

\def\Lr{{\rm Lr}}

The density is well defined and positive at zero chemical potential $\mu$,
so long as $0<\zdelta < 1$.  
The equation (\ref{2dboson.1}) has a real solution in this range
as long as $h>0$, i.e. as long as the gas is repulsive.    
This observation means that the bosonic instability described in
section IV can be cured by repulsive interactions. 
For the remainder of this section 
we set $\zmu =1$ and denote $\zdelta$ simply as $z$.   
Let us define $c$ as the following coefficient in the free energy:
\beq
\label{centralcboson}
\CF = -  \frac{ c  \pi   T^2  }{24}  
\eeq
This is analagous to the relativistic case in $1d$, where
one defines $\CF = - c \pi T^2 / 6$ where in that context
$c$ is the Virasoro central charge\cite{Cardy,Affleck}.    
In terms of the solution to the equation (\ref{2dboson.1}), 
$c$ is given in terms of Roger's dilogarithm: 
\beq
\label{cboson.2}
c(z) = \frac{ 6}{\pi^2}  ~ \Lr_2 (z) 
\eeq
where
\beq
\label{Rogers}
\Lr_2 (z) = \Li_2 (z) + {\textstyle \inv{2}} \log |z|  \log (1-z) 
\eeq
Interestingly,  the above formula for $c$ in terms of $\Lr_2$ 
is identical to the formulas that arise when one studies
the conformal limit of relativistic TBA systems in $1d$\cite{ZamoTBA,Klassen}. 

The function $\Lr_2$ is known to satisfy the following functional
relations\cite{Lewin}:
\barray
\label{lrfunc}
\Lr_2 (z) + \Lr_2 (1-z) &=& \zeta (2) = \frac{\pi^2}{6}
\\ \nonumber
\Lr_2 (z) + \Lr_2 ( -z/(1-z) ) &=& 0
\earray
Using the first relation,  one sees that $\Lr_2 (1/2) = \zeta(2)/2$. 
This, and the free fermion and free boson cases ($h=0$) were 
known to Euler\cite{Euler}.   Landen found more relations
as follows\cite{Landen}.   If $r$ is a root to the polynomial equation 
$z^2 + z =1$, then the above functional equations become linear
equations for $\Lr_2 (z)$ with argument
$z=r,-r, -1/r$ and $r^2$.   Let us chose 
$r=(\sqrt{5} -1 )/2$ which is the golden mean.  
When $z$ is positive,  the above relations 
imply that for special values of the coupling $h$, 
$c$ is a rational number:
\barray
\nonumber 
h &=& 0, ~~~~~~ z = 1,   ~~~~~~~ c= 1
\\ 
\label{rationalboson}
h&=& 1/2, ~~~z = r, ~~~~~~~ c=3/5 
\\ \nonumber
h &=& 1, ~~~~~~ z = 1/2 , ~~~~ c = 1/2 
\\ \nonumber
h &=& 2, ~~~~~~z  = r^2 , ~~~~~~ c=2/5
\earray
The above relations may be viewed as results in additive
number theory by virtue of eq. (\ref{nonrel.17}).

Since the pressure $p= - \CF$, the coefficient $c$ is
a measure of the pressure of the gas.    One observes
from eq. (\ref{rationalboson}) that as $h$ increases 
the pressure decreases.  This is expected on physical
grounds:  larger  $h$ means stronger repulsive interactions
so the gas is less dense.

\subsection{Fermionic case}

In this case the equation that determines $z$ (at zero chemical
potential) is 
\beq
\label{ferm2d.1}
z = ( 1 + z )^{-h}
\eeq
The density takes the form
\beq
\label{ferm2d.2}
n = \frac{T}{4\pi}    \log (1 + z ) 
\eeq
The free energy has the same form as in eq. (\ref{centralcboson}) 
where now
\beq
\label{ferm2d.4}
c  = - \frac{ 6}{\pi^2}  ~ \Lr_2 (-z)  
\eeq

Because of the Pauli exclusion principle for fermions, 
the fermionic gas is more stable than the bosonic one. 
First note that unlike the bosonic case, the eq. (\ref{ferm2d.1})
continues to have real solutions even if the interactions
of the gas are attractive with $h<0$.    Another important
feature is that whereas in the bosonic case $\Lr_2 (z)$ 
has a branch cut along the real $z$ axis from 1 to $\infty$, 
there are no branch points for $z$ along the negative axis.  
So for fermions, the free energy is well defined for 
any real $z<1$.  

The special rational points of the bosonic case also have
a fermionic version: 
\barray
\nonumber
h &=& -1, ~~~~~~~~ z = \infty , ~~~~~c=1 
\\ 
\label{fermrational}
h &=& -1/2, ~~~~~ z = 1/r , ~~~~ c= 3/5 
\\ \nonumber 
h &=& 0, ~~~~~~~~~~z = 1 , ~~~~~~~c=1/2
\\ \nonumber 
h &=& 1, ~~~~~~~~~~z = r, ~~~~~~~ c= 2/5
\earray

Note here also that increasing $h$ decreases the pressure
since the gas is either less attractive or more repulsive.

\subsection{Many mixed particles}

For far we have only considered a single bosonic or fermionic
particle.  It is straightforward to extend this to many types
of particles of mixed statistics.   Let $m_a$ and $s_a = \pm 1$
denote the mass and statistics of the $a$-th particle.   
In the two-body approximation, we consider the following
contribution to $\Fhatone$:
\beq
\label{many.1}
U = \frac{\beta}{2} \sum_{a,b} f_a (\kvec ) (\K_{ab} * f_b)(\kvec) 
\eeq
where $\K_{ab}$ is the scattering amplitude of $a$ with $b$ types
of particles.  
The saddle point equation now reads: 
\beq
\label{many.2}
\vep_a = \omega_a - \mu_a - \sum_b \K_{ab} * \inv{ e^{\beta \vep_b} -s_b }
\eeq
For a constant kernel $\K_{ab}$ the above equation has the solution:
\beq
\label{many.3}
\vep_a = \omega_a - \mu_a + T \delta_a
\eeq

In $2d$ the equation satisfied by $\delta_a$ is again algebraic: 
\beq
\label{many.4}
z_a = \prod_b ( 1- s_b z_b )^{h_{ab} s_b } 
\eeq
where 
$z_a = e^{-\delta_a}$ and $h_{ab}$ are dimensionless coupling
constants that parameterize the kernel $\K_{ab}$.   
The density $n_a$ of the a-th species is given by
\beq
\label{many.5}
n_a = - s_a \frac{m_a T}{2\pi} \log ( 1- s_a z_a ) 
\eeq
The total free energy is given by eq. (\ref{centralcboson}) 
with
\beq
\label{many.6}
c = 2 \sum_a  ~ m_a ~ c_{s_a} (z_a ) 
\eeq
where $c_+$  is the bosonic  expression (\ref{cboson.2})
and $c_-$ the fermionic one (\ref{ferm2d.4}).

The exists a vast number of examples where certain choices
of $m_a$, $s_a$, and $h_{ab}$ lead to rational $c$.   
For instance, one can translate known fermionic relativistic
systems in $1d$ to the present $2d$ non-relativistic context.
Some of the latter are known to be related to root systems of 
Lie algebras.  (See for instance \cite{Klassen}.)   
There are also many more examples that follow from results in
\cite{Lewin}, not all of which have $1d$ relativistic analog.  
For purposes of illustration,  consider a two-particle theory
with one boson and one fermion, of equal mass $1/2$,
and with $h_{ab} = 1 , ~~\forall a,b$.   This structure
suggests a supersymmetric theory.   The $z_1, z_2$
are solutions of 
\beq
\label{susy.1}
z_1 = \frac{1-z_1}{1+z_2}, ~~~~~z_2 = \frac{1-z_1}{1+z_2} 
\eeq
which implies $z_1 = z_2 = \sqrt{2} -1$.   It can be verified that this
theory has $c=3/4$.

\section{Quasi-periodic kernels and the Riemann hypothesis}

In order get into the critical strip of $\zeta (\nu)$ with
$0< \Re (\nu ) < 1$,  inspection of our work so far 
suggests that one could try  to analytically continue in the spacial
dimension $d$ to complex values.   However it would remain
unclear what complex $d$ actually means physically.    
A more physical approach  is to consider non-constant, quasi-periodic 
kernels in fixed  dimension $d$ an integer.     As we will show, 
$d=1$ is sufficient to cover the whole critical strip.  
Because of the previously discussed bosonic instabilities for $d<2$,
one must deal with a fermionic gas.   
In the next subsection we will simply hypothesize a certain kernel
$\K$ and work out its consequences.  In the  subsequent subsection
we will show how to obtain such a kernel from a second quantized
hamiltonian with a 2-body potential that is also quasi-periodic.

\subsection{Quasi-periodic kernel in $1d$}

We assume that the 2-body potential $\CV (\xvec , \xvec' )$ in
eq. (\ref{Ham.1}) is translationally invariant, i.e. 
it depends only on the difference $\xvec - \xvec' $.   
In momentum space this implies that the kernel also
depends only on the difference:
\beq
\label{quasi.1}
\K (\kvec ,  \kvec') = \K (\kvec - \kvec') 
\eeq
We also assume rotational invariance so that $\K$ depends only
on $k = |\kvec|$.   Let us suppose there exists a hamiltonian
which leads to the following kernel:
\beq
\label{quasi.2}
\K (k) = - \Re \( \gamma_\nu \, k^{2\nu -1} \)
\eeq
where $\nu$ is assumed to be a complex number and $\gamma_\nu$
is a constant.   Note that the lowest order constant
scattering length approximation vanishes as long 
as $\Re (\nu) > 1/2$. 

   The saddle point equation, with zero
chemical potential,  reads:
\beq
\label{quasi.2b}
\vep (\kvec ) = \omega_\kvec + \int \frac{d\kvec'}{2\pi} ~ 
\Re \( \gamma_\nu | \kvec - \kvec'|^{2\nu-1} \) ~
\inv{ e^{\beta \vep (\kvec' )} +1 }
\eeq
At low temperature, $\vep(k)$ can be approximated near
$k=0$,  where it continues to take the form 
eq. (\ref{nonrel.13}), with $\mu = 0$.  
The constant $\delta$ must satisfy the equation:
\beq
\label{quasi.3}
\delta = - \Re \[ T^{\nu-1} ~ h_\nu ~ \Li_\nu ( -\zdelta ) \]
\eeq
where $h_\nu = \gamma_\nu \Gamma(\nu) / 2\pi $ and as
before $\zdelta = e^{-\delta}$.  The density and free energy
however have the same expressions as in section IV specialized
to $d=1$: 
\barray
\label{quasi.4}
n &=& - ( T/4\pi)^{1/2} ~ \Li_{1/2} ( - \zdelta ) 
\\ \nonumber
\CF &=&  (T^3/4\pi)^{1/2} \( \Li_{3/2} (-\zdelta ) + \frac{\delta}{2} 
\Li_{1/2} ( -\zdelta ) \) 
\earray

There are some trivial $\delta=0$ solutions to eq. (\ref{quasi.3}) 
which arise when the prefactor $1-2^{1-\nu}$ in eq. (\ref{fermzeta})
equals zero.  
We can remove them by 
chosing  $\gamma_\nu = ((1 - 2^{1-\nu}) \Gamma(\nu) )^{-1}$.
We have divided by $\Gamma (\nu)$ in order to remove 
another trivial zero which can arise from $\Gamma (i \infty ) = 0$.    
The coupling constant then becomes:
\beq
\label{quasi.5}
h_\nu = - \frac{ 2^{(\nu -3)/2} }{2\pi \sinh \(  {\textstyle 
\frac{(1-\nu)\log(2)}{2} }  \)  }
\eeq

We can now give a clear meaning to a zero $\nu$ of the zeta function
$\zeta (\nu ) = 0$.     Because of the relation (\ref{fermzeta}), 
when $\nu$ is a zero, then a solution to eq. (\ref{quasi.3}) is
$\delta = 0$.  This solution exists for {\it any temperature T}.  
By plotting the left and right hand sides of eqn. (\ref{quasi.3}),
one sees that in general there is another solution at $\delta \neq 0$,
but the latter depends on $T$, and in fact for some $T$ this solution
disappears.  Thus for the generic 
 $\delta=0$ solution,    this means there are no corrections to the
free energy at arbitrary temperature $T$, 
i.e. the density and free energy are the same
as in a free (non-interacting) theory:
\barray
\label{free1d}
n &=& \( T/4\pi \)^{1/2} ( 1- \sqrt{2} ) \zeta (1/2) 
\\ \nonumber
\CF &=& - \( T^3/4\pi \)^{1/2} (1 - 1/\sqrt{2} ) \zeta (3/2) 
\earray
Note that the density is still positive since $\zeta(1/2)$ is negative:
$(1-\sqrt{2} )\zeta (1/2) \approx .60649$.   
The pressure is also positive: $(1-1/\sqrt{2}) \zeta (3/2) \approx
.76515$.

The Riemann hypothesis would then  follow from the line of reasoning:
If (i) The leading contributions to the pressure of the gas
are the ones calculated in this work, (ii) There is 
a hamiltonian that leads to the $\nu$-dependent quasi-periodic kernel 
(\ref{quasi.2}), and (iii) Non-zero interactions
necessarily modify the pressure of the gas over a range of temperatures,   
then $\zeta (\nu) $ can have no zeros. 
In the next subsection we address (ii).

\subsection{Real space potentials}

We now show that  there are indeed real space hamiltonians in $1d$ that
lead to the above quasi-periodic kernel.   
Let $v(\kvec )$ be the Fourier transform of the 2-body
potential: 
\beq
\label{real.1}
\CV (\xvec - \xvec' ) = \int \frac{d^d \kvec}{(2\pi)^d} ~ 
e^{i\kvec \cdot ( \xvec - \xvec' )} ~ v(\kvec ) 
\eeq
Using eq. (\ref{nonrel.8}) one finds
\beq
\label{real.3}
\K (\kvec ) = -\inv{4} \( v(\kvec ) + v(-\kvec ) + 2 s \,  v(0) \) 
\eeq
We have included the statistics parameter $s$ in order to 
point out some features.  

Consider the following potential: 
\beq
\label{real.4}
\CV (\xvec ) = \Re  \( \frac{b_\nu}{|\xvec|^{2\nu}} \)  
\eeq
where $b_\nu$ is a constant.  
If $b_\nu$ is real and $\nu = \sigma/2 + i \alpha$ then 
the potential is quasi-periodic in $\log |x|$:
\beq
\label{real.4b}
\CV (\xvec ) = \frac{ b_\nu \cos ( 2\alpha \log |\xvec | )}{|\xvec|^\sigma}
\eeq
This model is not known to be integrable, unlike the 
case of a delta-function potential, thus there is no
known exact thermodynamic Bethe ansatz.   The lowest order
corrections are the ones calculated in the last section.  

When the particles are fermionic with $s=-1$ the
$v(0)$ term in eq. (\ref{real.3}) precisely renders the kernel 
well behaved as $k\to 0$.  The kernel would be singular if
the particles were bosonic.  
The kernel is then given by the following integral:
\beq
\label{real.5}
\K ( k ) =   
\Re \( 2 b_\nu \int_0^\infty dx  ~x^{-2\nu } ~ \sin^2 ( k x /2 ) \)
\eeq
This integral is convergent if the following condition is
satisfied:
\beq
\label{restr}
  1/2 <  \Re (\nu ) < 3/2
\eeq
If the above condition is met, 
the kernel is the following: 
\beq
\label{real.6}
\K (k) = - \Re \[ b_\nu \, k^{2\nu -1} \sin (\pi \nu) 
\Gamma( {\textstyle 1-2\nu } ) \]
\eeq
Finally we can chose $b_\nu$ in order to obtain the kernel in 
eq. (\ref{quasi.5}).  Using the identity 
\beq
\label{real.7}
\sin (\pi \nu ) \Gamma({\textstyle 1-2\nu}) \Gamma (\nu) = \sqrt{\pi} 2^{-2\nu} 
\Gamma ({\textstyle 1/2 - \nu}  ) 
\eeq
this fixes $b_\nu$ to be:
\beq
\label{real.8}
b_\nu = - \frac{ 2^{(5\nu -3)/2} }{\sqrt{\pi} \Gamma \( {\textstyle \frac{1-2\nu}{2}} \) 
\sinh\( {\textstyle \frac{(1-\nu)\log(2)}{2} } \)  }
\eeq

The condition (\ref{restr}) is precisely what one needs for the
RH.    What is then the meaning of the known zeros at $\Re(\nu) = 1/2$?  
These are models with the kernel (\ref{quasi.2}) that give 
vanishing leading  corrections to the pressure.   The  calculations
of this section show that such a kernel does not arise in
a convergent manner from a real space potential since the
condition (\ref{restr}) is violated.    Note that 
the $\Gamma$ function in eq. (\ref{real.6}) develops
a pole at $\nu = 1/2$, which suggests that 
an additional low-energy regularization could still lead
to sensible models with the kernel (\ref{quasi.2}) 
that provide physical realizations of the Riemann zeros.

\section{Conclusions}

We have shown how the formalism developed in \cite{AndreSthermo} 
can lead to new results for the quantum statistical mechanics
of interacting gases of bosons and fermions.   
Our main results were summarized in the introduction.   

Clearly our most interesting result is the formulation of 
the Riemann hypothesis in the present context.    
We have essentially given a physical argument from which it follows.  
In order to develop this argument  into a rigorous mathematical
proof,  one mainly needs to give a firm foundation to the 
theoretical methods used here, namely the framework developed in
\cite{AndreSthermo}.    One also needs to understand 
more rigorously how the contributions considered in this paper
are the leading ones,  i.e. one needs clearer  control of the
approximations we have made.

\section{Acknowledgments}

I wish to thank Germ\'an Sierra for many discussions  on 
the Riemann hypothesis in connection with our work on
cyclic  renormalization group flows.

\section{Appendix}

In this appendix we explain how the $\nu \to 1-\nu$ duality
of Riemann's zeta function $\zeta (\nu)$ can be understood
as a special modular transformation in a Lorentz-invariant theory. 

Consider a free quantum field theory of massless bosonic particles
in $d+1$ spacetime dimenions with euclidean action
$S = \int d^{d+1} x  ~ (\d \phi)^2 $.   
The geometry of euclidean spacetime is take to be $S^1 \times R^d$
where the circumference of the circle $S^1$ is $\beta$.   We will
refer to the $S^1$ direction at ``$\hat{x}$''.  Endow the flat space
$R^d$ with a large but finite volume as follows.  Let us refer
to one the directions perpendicular to $\hat{x}$ as $\hat{y}$
with length $L$ and let the remaining $d-1$ directions have volume $A$. 

Let us first view the $\hat{x}$ direction as compactified euclidean time,
so that we are dealing with finite temperature $T=1/\beta$.   As
a quantum statistical mechanical system, the partition function 
in the limit $L, A \to \infty$ is
\beq
\label{ap.1}
Z = e^{-\beta V \, \CF (\beta )}
\eeq
where $V=L\cdot A$ and $\CF$ is the free energy density.  
Standard results give:
\beq
\label{ap.2}
\CF (\beta ) = \inv{\beta} \int \frac{d^d \kvec}{(2\pi)^d} ~
\log \( 1 - e^{-\beta k} \)
\eeq

The euclidean rotational symmetry allows one to view the above
system with time now along the $\hat{y}$ direction.  In $1d$,
interchanging the role of $\hat{x}$ and $\hat{y}$ is a special case
of a modular transformation of the torus.   In this version, 
the problem is a zero temperature quantum mechanical system with
a finite size $\beta$ in one direction,  and the total volume of
the system is $V'= \beta \cdot A$.   The quantum mechanical path
integral leads to 
\beq
\label{ap.3}
Z = e^{- L E_0 (A,\beta)}
\eeq
where $E_0$ is the ground state energy.  Let $\CE_0 = E_0/V'$ 
denote the ground state energy per volume.  Comparing the two
``channels'',  their equivalence requires $\CE_0 (\beta) = \CF (\beta)$.  
In this  finite-size channel,  the modes of the field in the
$\hat{x}$ direction are quantized with wave-vector $k_x = 2 \pi n/\beta$,
and the calculation of $\CE_0$ is as in the Casimir effect:
\beq
\label{ap.4}
\CE_0 = \inv{2\beta} \sum_{n \in \Zmath} 
\int \frac{ d^{d-1} \kvec}{ (2\pi)^{d-1} } 
\( \kvec^2 + (2\pi n /\beta)^2 \)^{1/2} 
\eeq

The free energy density $\CF$ can be calculated using $\int d^d \kvec = 
2\pi^{d/2} / \Gamma(d/2)$.  For $d>0$ the integral is convergent
and one finds
\beq
\label{ap.5}
\CF = -  \inv{\beta^{d+1}}  ~
\frac{ \Gamma (d+1) \zeta (d+1)}{2^{d-1} \pi^{d/2} \Gamma(d/2)d 
 } 
\eeq

For the Casimir energy, $\CE_0$ involves 
$\sum_{n \in \Zmath} |n|^d $ which must be regularized.  
As is usually done, let us regularize this as $2 \zeta (-d)$.  
Then:
\beq
\label{ap.6}
\CE_0 = - \inv{\beta^{d+1}} ~ \pi^{d/2} \Gamma(-d/2) \zeta (-d) 
\eeq

Define the function
\beq
\label{ap.7}
\xi (\nu) \equiv  \pi^{-\nu /2} \Gamma(\nu/2) \zeta (\nu) 
\eeq
Then the equality $\CE_0 = \CF$ requires the  identity:
\beq
\label{ap.8}
\xi (\nu ) = \xi (1-\nu) 
\eeq
The above relation is a known functional identity that 
can be proven using complex analysis. 
Thus we have demonstrated  that $\zeta$ function regularization
of the Casimir energy is consistent with a modular transformation
to the finite-temperature channel.   
On the other hand, our calculations can be viewed as a
proof of the identity (\ref{ap.8}) based on physical consistency.

\end{document}